\documentclass[%
 notitlepage
 amsmath,amssymb,
 aps,
]{revtex4-2}

\usepackage{amsmath}
\usepackage{bm}
\usepackage{graphicx}
\usepackage{dcolumn}
\usepackage{bm}
\usepackage{hyperref}
\usepackage{epstopdf}

\usepackage{color,soul}
\newcommand{\hla}[2][white]{{\sethlcolor{#1}\hl{#2}}}
\newcommand{\hlb}[2][white]{{\sethlcolor{#1}\hl{#2}}}
\newcommand{\hlc}[2][white]{{\sethlcolor{#1}\hl{#2}}}

\begin{document}
\title{A mesh-free framework for high-order simulations of viscoelastic flows in complex geometries}
\author{J. R. C. King}
\email{jack.king@manchester.ac.uk}
\affiliation{
School of Engineering, University of Manchester, UK
}%
\author{S. J. Lind}
\affiliation{
School of Engineering, Cardiff University, UK
}
\date{\today}

\begin{abstract}
The accurate and stable simulation of viscoelastic flows remains a significant computational challenge, exacerbated for flows in non-trivial and practical geometries. Here we present a new high-order meshless approach with variable resolution for the solution of viscoelastic flows across a range of Weissenberg number\hla{s}. Based on the \hlb{Local Anisotropic Basis Function Method (}LABFM\hlb{)} of King et al. \textit{J. Comput. Phys.} 415 (2020):109549, highly accurate viscoelastic flow solutions are found using Oldroyd B and PPT models for a range of two dimensional problems - including Kolmogorov flow, planar Poiseulle flow, and flow in a representative porous media geometry. Convergence rates up to $9^{th}$ order are shown. Three treatments for the conformation tensor evolution are investigated \hlc{for use in this new high-order meshless context} (direct integration, Cholesky decomposition, and log-conformation), with log-conformation providing consistently stable solutions across test cases, and direct integration yielding better accuracy for simpler unidirectional flows. The final test considers symmetry breaking in the porous media flow at moderate Weissenberg number, as a precursor to a future study of fully 3D high-fidelity simulations of elastic flow instabilities in complex geometries. The results herein demonstrate the potential of a viscoelastic flow solver that is both high-order (for accuracy) and meshless (for straightforward discretisation of non-trivial geometries including variable resolution). In the near-term, extension of this approach to three dimensional solutions promises to yield important insights into a range of viscoelastic flow problems, and especially the fundamental challenge of understanding elastic instabilities in practical settings. 
\end{abstract}
\maketitle
\section{Introduction}\label{sec:intro}
Despite decades of research effort, the determination of accurate viscoelastic flow solutions remains a key challenge in computational rheology. \hlb{There are a wide variety of techniques to improve numerical stability at higher levels of elasticity. Many early approaches saught to adjust the balance of elliptic and parabolic terms, with examples being elastic-viscous split-stress (EVSS) schemes~\mbox{\cite{beris_1984,RAJAGOPALAN_1990}}, adaptive viscoelastic stress splitting (AVSS)~\mbox{\cite{sun_1996,sun_1999}} and ``both sides diffusion'' (BSD)~\mbox{\cite{xue_2004,chen_2013}}. Whilst these can provide increase stability, especially near the limit of zero solvent viscosity, they do not provide stability in complex transient flows at high levels of elasticity. Another approach, commonly used in pseudo-spectral methods (e.g.~\mbox{\cite{morozov_2022,gillissen_2019}}) is to add some form of artificial diffusivity to the system. Whilst this can provide stability by imposing a limit on the smallest lengthscales of the flow, this is at the expense of accuracy, though justifications can be made by analogy with molecular diffusion. Perhaps the most significant development in the state-of-the-art has arisen from approaches which seek to transform the equations governing the evolution of the conformation tensor (or polymeric stress), such that (some of the) physical constraints are respected. Two significant examples of this are the log conformation formulation~\mbox{\cite{fattal_2004,fattal_2005}}, which guarantees the conformation tensor remains symmetric positive definite, and the Cholesky decomposition approach~\mbox{\cite{vaithianathan_2003}}.}

\hlb{The above developments have }improved capability greatly by increasing stability of simulations, especially as one enters higher Weissenberg, $Wi$, number regimes, where historically simulations would quickly fail approaching $Wi=\mathcal{O}\left(1\right)$. Whilst the stability of viscoelastic flow simulations has been greatly improved, \textit{accuracy} of these simulations is now a primary concern. Stable simulations, particularly those at higher $Wi$ number face considerable difficulty in attained accurate, converged solutions, due to extremely large elastic stress gradients (for example near solid boundaries) and/or the development of very thin transient elastic stress filaments often a precursor to 2D or 3D (visco-)elastic instability and potentially the onset of elastic or elasto-inertial turbulence ~\cite{steinberg_2021,dubief_2023}. The resolution required to resolve the elastic stresses is considerable – and as the $Wi$ number increases, obtaining fully converged solutions can become prohibitively expensive, with computational grid requirements dwarfing those required for the equivalent Newtonian turbulence simulation at the same Reynolds number. Given that understanding \textit{Newtonian} turbulence remains one of the great open challenges in fluid mechanics, the challenge facing computational rheology in this regard considerable.

In order to attain high degrees of solution accuracy in a practical time frame, high-order methods become essential. Spectral, spectral element, and $hp$ element methods have become established in computational rheology over the years~\cite{owens2002computational}, and there are many examples of their usage in solving a range of challenging viscoelastic flow problems and with a high degree of success (see for example,~\cite{pilitsis1989calculations,owens1993compatible,MOMENIMASULEH_2004,berti_2010,CLAUS_2013,KYNCH_2017}). In simple domain geometries (i.e. rectangular) spectral methods have few competitors: relatively fast and extremely accurate they have been used with great success to model higher $Wi$ number problems and fundamental flow studies in elastic turbulence (see~\cite{morozov_2022, Garg_2021}, for example). For more practical contexts however – namely complicated geometries perhaps resembling industrial processing/mixing devices where accurate flow solutions have broad utility and benefit (e.g. \cite{RAMSAY_2016}) – spectral methods are inapplicable. Spectral and p-finite element methods offer greater geometric flexibility, but, like mesh-based methods generally, constructing a mesh in a very complicated geometry that results in stable and converged solutions is particularly challenging and a significantly time-consuming task at pre-processing. Generally, element sizes and shapes have to sufficiently regular and well-distributed for accuracy and stability, which can be particularly difficult to achieve in very complicated geometries and makes an effective high-order adaptive or dynamic meshing scheme, e.g. for resolving thin transient elastic flow structures, particularly difficult to implement.

Meshless methods circumvent many of these challenges and make the process of domain discretisation much simpler in comparison. Meshless computational nodes have limited connectivity or requirements on topology and can often be scattered across a domain, then diffused or advected (by the flow or some transport velocity or otherwise) to improve node distribution. This may be done at pre-processing for Eulerian (fixed node) approaches or, for Lagrangian or Arbitrary-Lagrangian-Eulerian (ALE) simulations, during the simulation itself. Smoothed Particle Hydrodynamics is perhaps one of the most well know meshless methods and has been used to solve viscoelastic flow problems in the Lagrangian context for many years (see for example~\cite{ELLERO_2002,fang2006numerical,vq_ellero_2012,king_2021}). The computational nodes are simultaneously Lagrangian fluid elements, which can offer stability benefits in the context of viscoelastic flow simulation by effectively removing the advective term in the governing equations. There are also related methods, such as Dissipative Particle Dyanmics (DPD) and Smoothed Dissipative Particle Dynamics (SDPD), which are also subject to a concerted research effort in computational rheology~\cite{TENBOSCH_1999,PHANTHIEN_2018,litvinov2008,moreno2021arbitrary,NIETOSIMAVILLA_2022}, but these tend to apply on the physical length scales where the search for converged continuum solutions becomes less relevant.

While offering enviable geometric flexibility and stability, the issue with SPH and related approaches is accuracy. In its traditional form the SPH method is low order~\cite{quinlan}, and as discussed above, without high-order resolving power, the resolutions required for practical simulation become prohibitively expensive (especially as SPH is more computationally expensive than most grid-based methods at an equivalent resolution). The computational rheology community would therefore benefit from a method that is both meshless (to provide the improved geometric flexibility) and high-order (to provide the accuracy and resolving power). In recent years the authors and co-workers have been developing such a method; originally motivated by the need to create a high-order version of the SPH method \cite{lind_2016,nasar_2019}, \hlb{the Local Anisotropic Basis Function Method (}LABFM\hlb{)} has emerged as a generalised high-order meshless scheme \cite{king_2020,king_2022} with arbitrary orders of convergence possible (but 6th or 8th order spatial convergence is typical). Following the analysis of prototypical Newtonian flow cases in \cite{king_2020, king_2022}, LABFM has recently been extended to the study of combustion physics and flame-turbulence interactions in complex geometries in~\cite{king_2024}.

The potential of the LABFM approach as a geometrically flexible, multi-physics, high-order solver is such that \hlc{the aim and focus of} this manuscript \hlc{is} the extension of LABFM to the solution of viscoelastic fluid flow. Herein we demonstrate high-order solutions of viscoelastic flow in both simple and non-trivial geometries \hlc{with convergence rates of up to $9^{th}$ order possible. We also consider} three numerical approaches for the viscoelastic stresses (based on direct integration, Cholesky decomposition, and log conformation) \hlc{to assess suitability in this new high-order mesh-free context} for different test cases. The paper concludes with a preliminary study of a two-dimensional symmetry breaking elastic instability in a representative porous media geometry at moderate $Wi$, demonstrating the potential of high-order meshless schemes for the fundamental study of elastic instabilities and elastic turbulence in non-trivial geometries in the near future. It is hoped that in the longer term the method may serve as a practical tool for the computational analysis, optimisation and design of challenging industrial viscoelastic fluid processing, an activity which underpins healthcare product and foodstuff manufacturing, energy supply, and many other important industries worldwide.

The remainder of this paper is set out as follows. In Section~\ref{sec:ge} we introduce the governing equations, and their Cholesky and log-conformation formulations, and in Section~\ref{sec:ni} we describe the numerical implementation. Section~\ref{sec:nr} contains a set of numerical results providing validation for the model against two-dimensional Kolmogorov flow, Poiseuille flow, and flows past cylinders in a channel and representative porous media geometry. Section~\ref{sec:conc} is a summary of conclusions.

Before continuing further, we briefly comment on our notation. To avoid ambiguity, we use Einstein notation where possible, and, of the Latin characters, subscripts $i$, $j$ $k$, $l$, $n$ are reserved for this purpose, with repetition implying summation. Subscripts $a$ and $b$ are used for particle/node indexes. Bold fonts are used to refer to tensors in their entirety (e.g. $\bm{c}$) rather than individual components. The order of the spatial discretisation scheme is denoted by $m$.

\section{Governing Equations}\label{sec:ge}
In the present work we limit our focus to the two-dimensional problem. The governing equations for the density, momentum and conformation tensor (in Einstein notation) are
\begin{equation}\frac{\partial\rho}{\partial{t}}+\frac{\partial\rho{u}_{k}}{\partial{x}_{k}}=0,\label{eq:mass}\end{equation}
\begin{equation}\frac{\partial\rho{u}_{i}}{\partial{t}}+\frac{\partial\rho{u}_{i}u_{k}}{\partial{x}_{k}}=-\frac{\partial{p}}{\partial{x}_{i}}+\beta\eta\frac{\partial^{2}u_{i}}{\partial{x}_{k}\partial{x}_{k}}+\frac{\left(1-\beta\right)\eta}{\lambda}\frac{\partial{c}_{ki}}{\partial{x}_{k}}+\rho{f}_{i},\label{eq:mom}\end{equation}
\begin{equation}\frac{\partial{c}_{ij}}{\partial{t}}+u_{k}\frac{\partial{c}_{ij}}{\partial{x}_{k}}-\frac{\partial{u}_{i}}{\partial{x}_{k}}c_{kj}-\frac{\partial{u}_{j}}{\partial{x}_{k}}c_{ik}=-\frac{\left(c_{ij}-\delta_{ij}\right)}{\lambda}\left[1-2\varepsilon+\varepsilon{c}_{kk}\right],\label{eq:c}\end{equation}
in which $x_{i}$ is the $i$-th coordinate ($x_{i}=x,y$ for $i=1,2$), $u_{i}$ is the $i$-th component of velocity ($u,v$ for $i=1,2$), $\rho$ is the density, $p$ the pressure, $c_{ij}$ the $ij$-th element of the conformation tensor, $f_{i}$ is a body force, $\lambda$ the polymer relaxation time, $\eta$ the total viscosity, $\beta$ the ratio of solvent to total viscosity, and $\varepsilon$ is a non-linearity parameter. The system is closed with an isothermal equation of state $p=c^{2}_{s}\left(\rho-\rho_{0}\right)$, where $\rho_{0}$ is a reference density and $c_{s}$ is the sound speed. Taking $U$ and $L$ to be characteristic velocity and length scales, and $T=L/U$ the characteristic time scale, the governing dimensionless parameters are:
\begin{equation}Re=\frac{\rho_{0}UL}{\eta};\qquad{Wi}=\frac{\lambda{U}}{L};\qquad{Ma}=\frac{U}{c_{s}};\end{equation}
the viscosity ratio $\beta$, and the PTT nonlinearity parameter $\varepsilon$. Additional terms which might arise in~\eqref{eq:mom} and~\eqref{eq:c} due to compressibility (see e.g.~\cite{lind2010numerical,mackay2019derivation}) are neglected as we operate near the incompressible limit, with Mach number $Ma<0.05$ for all considered cases. In this paper, three different formulations are investigated for the conformation tensor evolution equation~\eqref{eq:c} to inform optimal use in the high-order meshless context across test cases. In particular, we employ direct numerical integration of~\eqref{eq:c}, Cholesky Decomposition~\cite{vaithianathan_2003}, and log-conformation ~\cite{fattal_2004,fattal_2005}, summaries of which are provided in the sections below. 

\subsection{Cholesky Decomposition}
First considered in the context of viscoelastic numerical simulation by~\cite{vaithianathan_2003}, the Cholesky decomposition of the conformation tensor offers a convenient way to maintain symmetric positive definiteness. Consider the Cholesky decomposition of the 2-D conformation tensor, viz., 
\begin{equation}\bm{c}=\begin{bmatrix}l_{11}^{2}&l_{11}l_{12}\\l_{11}l_{12}&l_{12}^{2}+l_{22}^{2}\end{bmatrix},\end{equation}
with $l_{ij}$ denoting the components of the lower triangular matrix $\bm{L}$, such that $\bm{c}=\bm{L}\bm{L}^T$.
\hlb{Defining $S_{ij}$ as the right hand side of~\mbox{\eqref{eq:c}}}
\begin{equation}S_{ij}=-\frac{\left(c_{ij}-\delta_{ij}\right)}{\lambda}\left[1-2\varepsilon+\varepsilon{c}_{kk}\right],\end{equation}
then the equations for the evolution of the Cholesky decomposition components are:
\begin{subequations}\begin{align}
\frac{\partial{l}_{11}}{\partial{t}}+u_{k}\frac{\partial{l}_{11}}{\partial{x}_{k}}&=\frac{\partial{u}}{\partial{x}}l_{11}+\frac{\partial{u}}{\partial{y}}l_{12}+\frac{S_{11}}{2l_{11}}\label{eq:l11}\\
\frac{\partial{l}_{12}}{\partial{t}}+u_{k}\frac{\partial{l}_{12}}{\partial{x}_{k}}&=\frac{\partial{v}}{\partial{y}}l_{12}+\frac{\partial{u}}{\partial{y}}\frac{l_{22}^{2}}{l_{11}}+\frac{\partial{v}}{\partial{x}}l_{11}+\frac{S_{12}}{l_{11}}-\frac{l_{12}}{2l_{11}^{2}}S_{11}\label{eq:l12}\\
\frac{\partial{l}_{22}}{\partial{t}}+u_{k}\frac{\partial{l}_{22}}{\partial{x}_{k}}&=\frac{\partial{v}}{\partial{y}}l_{22}-\frac{\partial{u}}{\partial{y}}\frac{l_{12}l_{22}}{l_{11}}+\frac{S_{22}}{2l_{22}}-\frac{S_{12}l_{12}}{l_{22}l_{11}}+\frac{l_{12}^{2}S_{11}}{2l_{11}^{2}l_{22}}\label{eq:l22}\end{align}\end{subequations}
Equations \eqref{eq:l11}-\eqref{eq:l22} can then be discretised in space and integrated in time as detailed in Section \ref{sec:ni}. \hlc{We refer to this formulation as CH. It is common practice to evolve the natural logarithms of $l_{11}$ and $l_{22}$. The corresponding evolution equations can be obtained by simply dividing~\mbox{\eqref{eq:l11}} by $l_{11}$ and~\mbox{\eqref{eq:l22}} by $l_{22}$ (with~\mbox{\eqref{eq:l12}} unchanged). We refer to this formulation as Cholesky-log, or CH-L}.

\subsection{Log-conformation}

The log-conformation formulation employed directly follows that of~\cite{fattal_2004,fattal_2005}, to which readers may refer for further details. \hlb{We define the diagonalisation of the conformation tensor as}
\begin{equation}c_{ij}=R_{ik}\Lambda_{kl}R_{jl},\end{equation}
\hlb{in which ${R}_{ij}$ contains the eigenvectors of the conformation tensor, and the diagonal matrix ${\Lambda}_{ij}$ contains the eigenvalues. We then }denote the log-conformation tensor as
\begin{equation}\Psi_{ij}=R_{ik}\ln\Lambda_{kl}R_{jl},\end{equation}
\hlb{where the logarithm is applied independently to each diagonal element ${\Lambda}_{ii}$.} The log-conformation tensor is then evolved according to
\begin{equation}\frac{\partial\Psi_{ij}}{\partial{t}}+u_{k}\frac{\partial\Psi_{ij}}{\partial{x}_{k}}=\Omega_{ik}\Psi_{kj}-\Psi_{ik}\Omega_{kj}+2{B}_{ij}-\frac{1}{\lambda}\left(1-2\varepsilon+\varepsilon{c}_{kk}\right)\left\{c^{-1}\right\}_{ij}\left({c}_{ij}-\delta_{ij}\right)\label{eq:lc},\end{equation}
where
\begin{equation}\Omega_{ij}=R_{ik}\left(\omega_{kn}\left(1-\delta_{nl}\right)\right)R_{jl},\end{equation}
and
\begin{equation}B_{ij}=R_{ik}\left(m_{kn}\delta_{nl}\right)R_{jl},\end{equation}
in which
\begin{equation}\omega_{ij}=\left(\Lambda_{j}m_{ij}+\Lambda_{i}m_{ji}\right)/\left(\Lambda_{j}-\Lambda_{i}\right)\end{equation}
and
\begin{equation}m_{ij}=R_{ki}\frac{\partial{u}_{k}}{\partial{x}_{l}}R_{lj}.\end{equation}

The subsequent spatial and temporal discretisation of the log-conformation scheme is described in Section \ref{sec:ni}. As mentioned, both the Choleksy decompostion and log-conformation approaches will be compared with a scheme that directly integrates the conformation tensor evolution equation \eqref{eq:c}, with no explicit constraints given on tensor positive definiteness. The aim will be to compare the three schemes across different flow test cases to determine optimal usage in the high-order meshless framework.  

\section{Numerical implementation}\label{sec:ni}

For all three formulations, the numerical implementation closely follows that described in~\cite{king_2022} for isothermal Newtonian flows and~\cite{king_2024} for turbulent reacting flows, with spatial discretisation based on the Local Anisotropic Basis Function Method (LABFM), time integration via an explicit Runge-Kutta scheme, and acceleration via OpenMP and MPI, with non-uniform structured block domain decomposition.

\subsection{Spatial discretisation}

The spatial discretisation is based on LABFM, which has been detailed and extensively analysed in~\cite{king_2020,king_2022}, and we refer the reader to these works for a complete description\hla{, but provide sufficient details here to reproduce the method}. Briefly summarising, the domain is discretised with a point cloud of $N$ nodes, unstructured internally, and with local structure near boundaries. \hlb{The node-set is fixed in space; i.e. the nodes do not move during the simulation. Along wall, inflow and outflow boundaries, nodes are distributed uniformly. Near boundaries, an additional $4$ rows of uniformly distributed nodes are arranged along boundary normals originating at nodes on boundaries. These are used to construct one-sided difference operators on the boundaries. Internally, the node distribution is generated using the propagating front algorithm of~\mbox{\cite{fornberg_2015a}}. The generation of the node distribution is akin to a mesh-generation procedure for unstructured mesh-based methods, and we do not repeat details of the algorithm here, which has been described in~\mbox{\cite{king_2022,king_2024}}. To ensure repeatability of our work, for each case herein, node distributions are available on request.}

Each node $a$ has an associated resolution $s_{a}$, which corresponds to the local average distance between nodes. The resolution need not be uniform, and $s_{a}$ may vary with $x$ and $y$. Each node also has an associated computational stencil length-scale $h_{a}$, which again may vary with $x$ and $y$. The ratio $s_{a}/h_{a}$ is approximately uniform, with some variation due to the stencil optimisation procedure described in~\cite{king_2022}, which ensures the method uses stencils marginally larger than the smallest stable stencil. Each node holds the evolved variables $\rho_{a}$, $\rho{u}_{i,a}$, and either the components of $\bm{c}$, the components of $\bm{\Psi}$, or the components of the $\bm{L}$. The governing equations are solved on the set of $N$ nodes. 

\begin{figure}
    \includegraphics[width=0.49\textwidth]{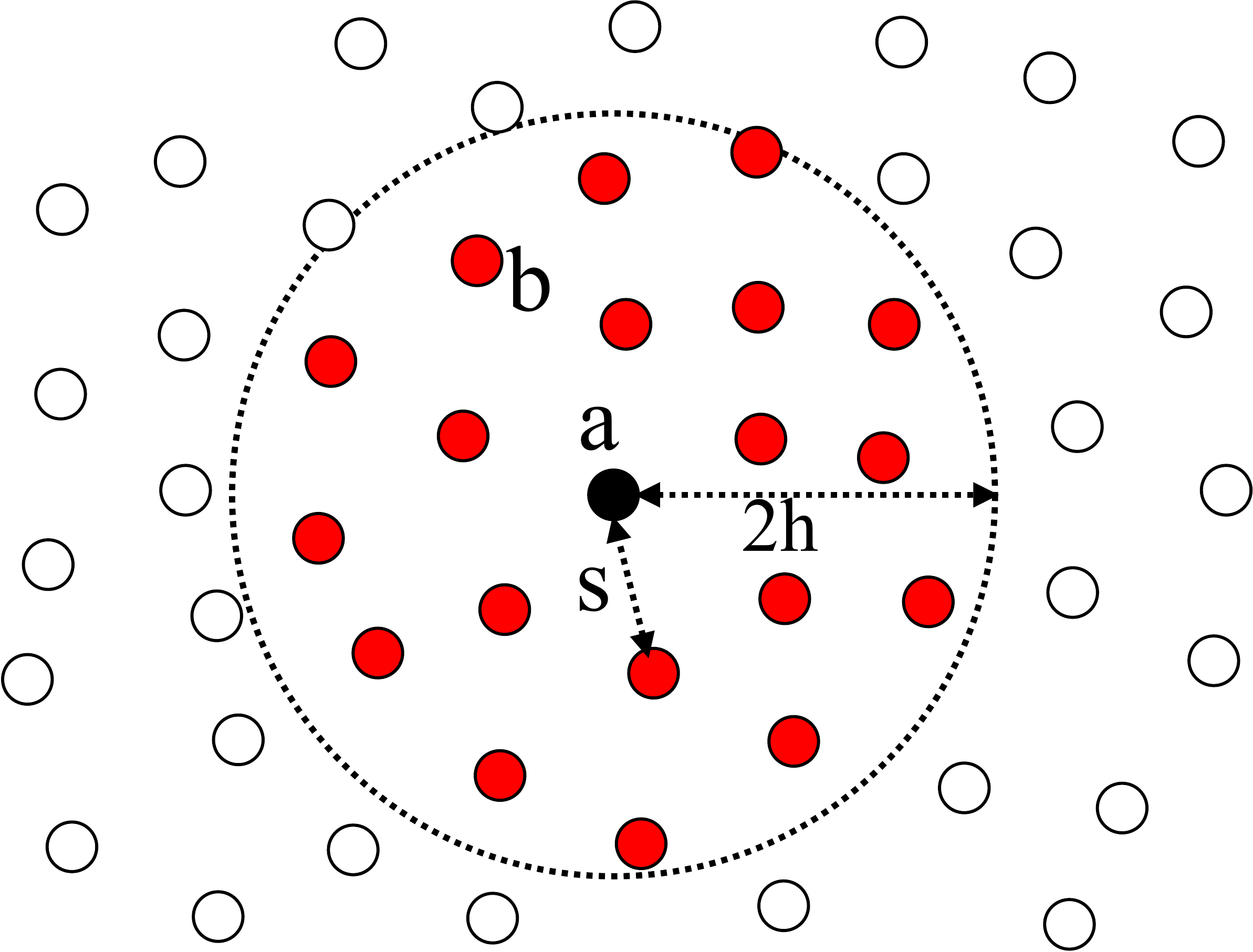}
    \caption{\hla{A schematic of the computational stencil.}\label{fig:stencil}}    
\end{figure}

The difference between properties at two nodes is denoted $\left(\cdot\right)_{ba}=\left(\cdot\right)_{b}-\left(\cdot\right)_{a}$. The computational stencil for each node $a$ is denoted $\mathcal{N}_{a}$, and is constructed to contain all nodes $b$ such that \hla{$r^{2}_{ba}={x}_{ba}^{2}+y_{ba}^{2}\le4h_{a}^{2}$}. The node distribution is generated using a variation of the propagating front algorithm of~\cite{fornberg_2015a}, following~\cite{king_2022}. \hla{A schematic of the computational stencil is shown in Figure~\mbox{\ref{fig:stencil}}.}

\hla{In the following, we use standard multi-index notation with index (e.g.) $\alpha$ representing the ordered pair $\left(\alpha_{1},\alpha_{2}\right)$, with $\left\lvert{\alpha}\right\rvert=\alpha_{1}+\alpha_{2}$. For clarity, as is standard in multi-index notation $\alpha!=\alpha_{1}!\alpha_{2}!$, $x^{\alpha}=x_{1}^{\alpha_{1}}x_{2}^{\alpha_{2}}$, and the partial derivative}
\begin{equation}\partial^{\alpha}=\frac{\partial^{\alpha_{1}}}{\partial{x}_{1}}\frac{\partial^{\alpha_{2}}}{\partial{x}_{2}}.\end{equation}
\hla{In LABFM, a}ll spatial derivative operators take the form
\begin{equation}L^{\gamma}_{a}\left(\cdot\right)=\displaystyle\sum_{b\in\mathcal{N}_{a}}\left(\cdot\right)_{ba}w^{\gamma}_{ba},\label{eq:general_do}\end{equation}
\hla{where $\gamma$ is a multi-index which identifies the derivative being approximated by~\mbox{\eqref{eq:general_do}}}, and $w^{\gamma}_{ba}$ are a set of inter-node weights for the operator. \hla{To evaluate $w^{\gamma}_{ba}$, we first define the vector of monomials $\bm{X}_{ba}$ element-wise, with the element corresponding to multi-index $\alpha$}
\begin{equation}X_{ba}=\frac{x_{ba}^{\alpha}}{\alpha!},\end{equation}
\hla{and a vector of anisotropic basis functions $\bm{W}_{ba}=\bm{W}\left(\bm{x}_{ba}\right)$, with the element corresponding to multi-index $\alpha$ given by}
\begin{equation}W_{ba}=\frac{\psi\left({r}_{ba}/h_{a}\right)}{\sqrt{2^{\left\lvert{\alpha}\right\rvert}}}H_{\alpha}\left(\frac{x_{ba}}{h_{a}\sqrt{2}}\right),\end{equation}
\hla{where }
\begin{equation}H_{\alpha}\left(x\right)=H_{\alpha_{1}}\left(x_{1}\right)H_{\alpha_{2}}\left(x_{2}\right)\end{equation}
\hla{are bi-variate Hermite polynomials of the physicists kind, and the radial basis function (RBF) $\psi$ is a Wendland C2 kernel~\mbox{\cite{dehnen_aly}}.} \hla{The weights $w_{ba}^{\gamma}$ in~\mbox{\eqref{eq:general_do}} are constructed as}
\begin{equation}w^{\beta}_{ba}=\bm{W}_{ba}\cdot\bm{\Psi}^{\beta},\end{equation}
\hla{with $\bm{\Psi}^{\gamma}$ a vector to be determined.} \hla{To determine $\bm{\Psi}^{\gamma}$ we construct and solve the linear system}
\begin{equation}\left[\displaystyle\sum_{b\in\mathcal{N}_{a}}\bm{X}_{ba}\otimes\bm{W}_{ba}\right]\cdot\bm{\Psi}^{\gamma}=\bm{C}_{\gamma},\label{eq:lsys}\end{equation}
\hla{in which $\bm{C}_{\gamma}$ is a unit vector defined element-wise as $C_{\gamma}^{\alpha}=\delta_{\alpha\beta}$, with $\delta_{\alpha\gamma}$ the Dirac-delta function.} \hla{The consistency of the operator~\mbox{\eqref{eq:general_do}} is then determined by the size of the linear system~\mbox{\eqref{eq:lsys}}. If we include the first $M=\frac{m^{2}+3m}{2}$ terms (i.e. up to order $\lvert\alpha\rvert\le{m}$) then the operator has polynomial consistency of order $m$. Consequently, first derivative operators converge with $s^{m}$, and second derivatives with $s^{m-1}$.}
\hla{With our nodes fixed in space, as a preprocessing step, for each node we construct and solve the linear system~\mbox{\eqref{eq:lsys}} to obtain $\bm{\Psi}^{\gamma}$}\hlb{, for $\gamma$ corresponding to both first spatial derivatives, and the Laplacian,}\hla{ which we then use to calculate and store $w_{ba}^{\gamma}$ in~\mbox{\eqref{eq:general_do}}.} \hlb{The derivatives appearing in~\mbox{\eqref{eq:mass}},~\mbox{\eqref{eq:mom}}, and either~\mbox{\eqref{eq:c}},~\mbox{\eqref{eq:l11}} to~\mbox{\eqref{eq:l22}}, or~\mbox{\eqref{eq:lc}} are approximated using~\mbox{\eqref{eq:general_do}}}.

\hlb{We highlight here that the consistency correction procedure described above removes the \emph{discretisation error limit} which causes a saturation of convergence in low-order mesh-free methods, such as SPH (see e.g.~\mbox{\cite{quinlan}} for a discussion). In such methods, this limit is caused by the fact that the kernel doesn't account for the node distribution, and the derivation of the method assumes the equivalence of the integral of the kernel over its support with the sum of the kernel over the particles in a stencil. No such assumptions are made in the present method, and the operators converge until an error limit dictated by the accuracy with which~\mbox{\eqref{eq:lsys}} can be solved, typically $\mathcal{O}\left(10^{-12}\right)$ for order $m=8$. For in-depth analysis of the LABFM discretisation, we refer the reader to~\mbox{\cite{king_2020,king_2022}}.}

The order of the spatial discretisation can be specified between $m=4$ and $m=10$, and although there is capability for this to be spatially (or temporally) varying, in this work we set $m=8$ uniformly away from boundaries (except where explicitly stated in our investigations of the effects of changing $m$). Whilst a larger value of $m$ gives greater accuracy for a given resolution, it also requires a larger stencil (larger $h_{a}/s_{a}$), and hence incurs greater computational expense for a given resolution. \hlb{As in central finite difference schemes, the cost scales with the product of the stencil size and the number of points $\mathcal{N}N$, where $\mathcal{N}$ is the average number of nodes in a computational stencil. In the present implementation, parallelised with MPI, scaling results have shown parallel efficiency above $96\%$ between $4$ and $1024$ cores. The method exhibits the same scaling performance as our related work in~\mbox{\cite{king_2024}}.} A value of $m=8$ provides a good compromise between accuracy and computational cost. At non-periodic boundaries, the consistency of the LABFM reconstruction is smoothly reduced to $m=4$. The stencil scale $h_{a}$ is initialised to $h_{a}=2.7s_{a}$ in the bulk of the domain, and $h_{a}=2.4s_{a}$ near boundaries (choices informed through experience as being large enough to ensure stability with $m=8$). In bulk of the domain, the stencil scale is then reduced following the optimisation procedure described in~\cite{king_2022}. This has the effect of both reducing computational costs, and increasing the resolving power of LABFM, and accordingly $h_{a}$ takes a value slightly larger than the smallest value for which the discretisation remains stable. \hla{Note that as with high-order central finite differences, or pseudo-spectral methods, no upwinding is used in the present scheme.}

\subsection{Temporal discretisation}

For the direct integration of $\bm{c}$ (hereafter referred to as DI), we evolve only $3$ of the $4$ (in two-dimensions) components of $\bm{c}$, and impose symmetry. There is nothing in this formulation to ensure $\bm{c}$ remains positive definite. Both the log-conformation (referred to as LC) and Cholesky decomposition (referred to as CH) formulations ensure $\bm{c}$ is symmetric positive definite by construction.

Time integration is by explicit third order Runge-Kutta scheme. We use a low-storage four-stage Runge-Kutta scheme with an embedded second order error estimator, with the designation RK3(2)4[2R+]C in the classification system of~\cite{kennedy_2000}. The value of the time-step is controlled using a Proportional-Integral-Derivative (PID) controller as described in~\cite{king_2024}, which ensures errors due to time-integration remain below $10^{-4}$. In addition to the PID controller, we impose an upper limit on the time step with 
\begin{equation}
\delta{t}\le\delta{t}_{max}=\max\left(\delta{t}_{cfl},\delta{t}_{visc}\right)
\end{equation}
where $\delta{t}_{cfl}=\min\left(s/\left(u+c_{s}\right)\right)$ and $\delta{t}_{visc}=\min\left(C_{visc}\eta{s}^{2}\right)$ denote the time steps due to the CFL and viscous diffusion constraints, respectively (with the $\min$ taken over the entire domain). We set the coefficient $C_{visc}$ to the limiting value $C_{visc}=1$, with the PID controller reducing the time-step size as necessary to keep time-integration errors bounded. 

\hla{We note here that because we use an explicit time integration scheme, at small $Re$, $\delta{t}\propto{Re}$, making simulations in the creeping flow limit prohibitively expensive. An implicit time-integration scheme could be implemented with the present spatial discretisation scheme. This would remove the viscous constraint on the time-step, at the expense of the solution of a large sparse linear system every step. Implicit schemes have been implemented in Smoothed Particle Hydrodynamics~\mbox{\cite{king_2021,king_2023a}}, which although low-order and Lagrangian, has effectively the same stencils as the present method, including on massively parallel architectures~\mbox{\cite{guo_2018}}, and for high-order variants on multiple GPUs~\mbox{\cite{oconnor_2022}} In Lagrangian schemes the linear system must be reconstructed each time-step, incurring considerable cost. In the present method, where the nodes are fixed, the linear system would need to be built only once. The development of an implicit, incompressible formulation of our numerical framework is planned, but beyond the scope of the present work.}

\hla{In common with other high-order collocated methods, the discretisation admits solutions with energy at the wavenumber of the resolution, and some form of de-aliasing or filtering is required. In the present work, the solution is dealiased every time-step by a high-order filter, as is commonly used for high-order central finite differences applied to compressible flows. For a field $\phi$, the filtering procedure is defined as}
\begin{equation}\hat\phi=\left(1+\mathcal{F}_{a}\right)\phi,\label{eq:filt}\end{equation}
\hla{in which the operator $\mathcal{F}_{i}$ is}
\begin{equation}\mathcal{F}_{a}\left(\phi\right)=\kappa_{m,a}\displaystyle\sum_{b\in\mathcal{N}_{a}}\phi_{ba}w_{ba}^{\gamma_{m}},\end{equation}
\hla{
where $\gamma_{m}$ is a multi-index such that the operator using weights $w_{ba}^{\gamma_{m}}$ approximates $\nabla^{m}$, and the pre-factor $\kappa_{m,a}$ is calculated in a pre-processing step as}
\begin{equation}\kappa_{m,a}=\frac{2}{3}\left\{\displaystyle\sum_{b\in\mathcal{N}_{a}}\left[1-\cos\left(\frac{3\pi{x}_{ba}}{2s_{a}}\right)\cos\left(\frac{3\pi{y}_{ba}}{2s_{a}}\right)\right]w_{ba}^{\gamma_{m}}\right\}^{-1}.\end{equation}
\hla{By construction, $\kappa_{m,a}$ ensures that the amplitude responses of the filter at a wavenumber two thirds of the Nyquist wavenumber (defined by the resolution $s$) is equal to $1/3$. Further details of the procedure can be found in~\mbox{\cite{king_2022,king_2024}}.} \hlb{The filter primarily acts on large wavenumbers, and has little effect on wavenumbers which are small relative to the Nyquist wavenumber of the discretisation ($\pi/s$). Provided the flow field is sufficiently resolved that the physical information in the solution lies at small wavenumbers (relative to $\pi/s$), the effect of the filter on the physical solution is negligible.} \hla{In the present work, the filter~\mbox{\eqref{eq:filt}} is applied every time step to the density and momentum fields, to either the components of $\bm{c}$ (for DI), the components of $\bm{\Psi}$ (LC) or the components of $\bm{L}$ (CH).}

\subsection{Boundary treatment}
Throughout this work, the computational domain is discretised with a strip of uniformly arranged nodes near solid boundaries, and an unstructured node-set internally. The discretisation procedure is the same as that used in~\cite{king_2022} and~\cite{king_2024}, to which we refer the interested reader for details. As in those works, numerical boundary conditions for no-slip walls are implemented via the Navier-Stokes Characteristic boundary condition formalism following ~\cite{sutherland_2003}, the implementation of which directly follows~\cite{king_2022}. For the conformation tensor (or its decompositions), the hyperbolic term is zero on solid boundaries, and requires no additional treatment. The upper-convected terms may be directly evaluated using the values of $\bm{c}$ (or its decompositions) on the boundary, and the local velocity gradients (evaluated using one-sided derivatives as detailed in~\cite{king_2022}).

\section{Numerical results}\label{sec:nr}

For the following, we reiterate use of the acronyms DI, CH and LC to indicate direct integration of the conformation tensor, Cholesky decomposition, and the log-conformation formulation, respectively. Where a figure provides a comparison of these formulations, we use red, black and blue lines for DI, LC and CH respectively.

\subsection{Kolmogorov flow}

\begin{figure}
    \includegraphics[width=0.49\textwidth]{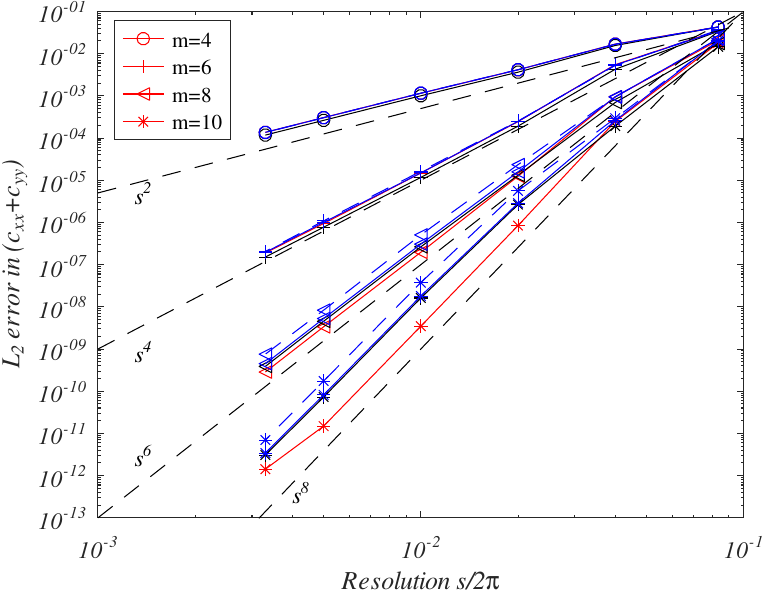}
    \includegraphics[width=0.49\textwidth]{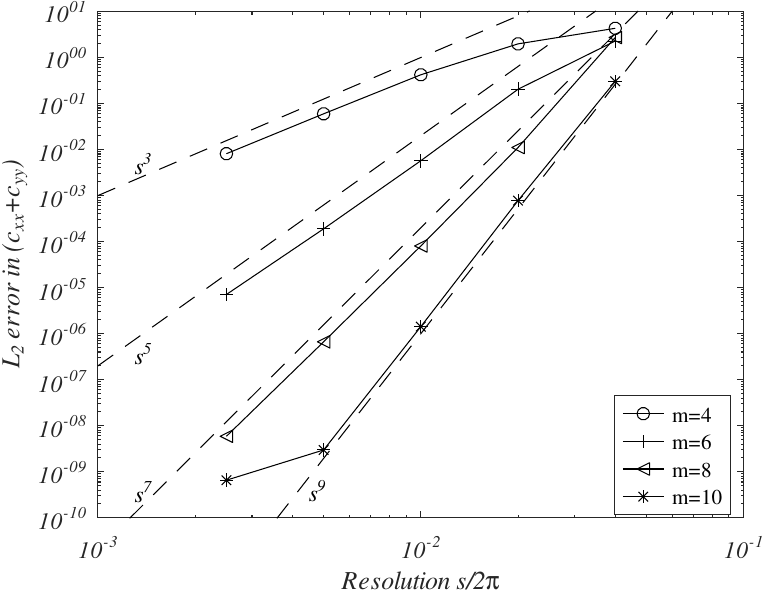}    
    \caption{Kolmogorov flow. Variation of $L_{2}$ error in steady state conformation tensor trace $c_{xx}+c_{yy}$, for different orders $m$ of the numerical discretisation scheme. The left panel shows the errors for all \hlc{formulations DI (red lines), CH (solid blue lines), CH-L (dashed blue lines)} and LC (black lines), for $Re=1$ and $Wi=0.1$. The right panel shows the errors for DI only, for $Re=10$ and $Wi=1$.\label{fig:klmgrv}}    
\end{figure}

Our first test is two-dimensional Kolmogorov flow. Although a simple geometry and one for which pseudo-spectral methods are well suited, the flow has an analytic solution in the steady state, and given the absence of boundaries, is a good test of the convergence of our method. The domain is a doubly-periodic square with side length $2\pi$. A forcing term of $f_{x}=\frac{4}{Re}\cos\left(2y\right)$ and $f_{y}=0$ is added to the right hand side of the momentum equation. The analytic solution in the steady state is $u=\cos\left(2y\right)$, $v=0$, $\rho=1$ and
\begin{equation}\bm{c}=\begin{bmatrix}1+8Wi^{2}\sin^{2}\left(2y\right)&-2Wi\sin\left(2y\right)\\-2Wi\sin\left(2y\right)& c_{yy}=1\end{bmatrix}.\end{equation}
We first set $Re=1$, $\beta=0.5$, $\varepsilon=0$, $Ma=0.05$. To reduce the costs of reaching a steady state, we use a small value of $Wi=0.1$. Note that the purpose of this test is not to demonstrate the ability of the method to reach larger $Wi$, but to assess the accuracy of the numerical scheme. The convergence rates observed would be unaffected if we set $Wi=1$, but the simulation would take longer to reach a steady state. We vary the order $m$ of the discretisation, with $m\in\left[4,6,8,10\right]$. The left panel of Figure~\ref{fig:klmgrv} shows the convergence in the trace of the conformation tensor for all three formulations for $Re=1$ and $Wi=0.1$. The errors scale with $s^{m-2}$. At higher $m$, DI is slightly more accurate than LC and CH, \hlc{and CH-L is least accurate, }but at lower $m$ \hlc{the errors for all formulations are} almost identical.

For lower $Re$, the time-step selection criteria is such that we have $\delta{t}\propto{s}^{2}$, and, by considering a steady state at a fixed time $t_{end}$, the total number of time-steps to reach $t_{end}$ is $N_{steps}\propto\delta{t}^{-1}\propto{s}^{-2}$. \hlb{The total error at $t_{end}$ is given by the spatial error, dominated by terms $\propto{s}^{m}$ in the present case, multiplied by the number of time steps, resulting in scaling of $s^{m-2}$,} as shown in the convergence rates of the left panel of Figure~\ref{fig:klmgrv}. The right panel of Figure~\ref{fig:klmgrv} shows convergence in $tr{\bm{c}}$ at larger $Re$ (and $Wi$, with $Re=10$ and $Wi=1$), but presenting the DI formulation only. Note that as the time step is now proportional to $s$ (rather than $s^2$), given the larger $Re$, convergence rates tend to follow order $m-1$, as observed in~\cite{king_2022}. For the right panel of Figure~\ref{fig:klmgrv} errors are taken at dimensionless time $t=40/\pi$, after the steady state is reached, but before the growth of any instability. The magnitudes of the errors are larger in this case because $Wi$ is larger (with similar behaviour seen in the Poiseuille flow case to follow in Section \ref{sec:pois}). For a fixed $Re$, increasing $Wi$ will increase the error magnitude, but leave the rate of convergence unchanged.

For both panels in Figure~\ref{fig:klmgrv} we see slightly lower convergence rates at very coarse resolutions because the wavelengths on which the high-order filters act are closer to those of the base solution. This behaviour has been observed and discussed in~\cite{king_2022}, and further in the context of three-dimensional turbulence simulations in~\cite{king_2024}. If filtering for the conformation tensor equation is removed, this \emph{does not} affect the order of convergence, but does reduce the magnitude of the errors by $\mathcal{O}\left(10^{3}\right)$. However, at coarse resolutions, filtering is essential for stability with $\bm{c}$ becoming unbounded in the absence of filtering for all three formulations (DI, CH, LC).

\subsection{Poiseuille Flow}\label{sec:pois}

Poiseuille flow is an important and classical flow test admitting analytical solutions for Oldroyd B fluids in the unsteady transient start-up flow as well as at steady state, and hence provides a good test of method accuracy. For all Poiseuille flow tests considered, the domain is a unit square, periodic laterally with no-slip wall boundaries at the top and bottom. The flow is driven by a constant and uniform body force such that the channel centreline velocity in the steady state is unity.

\subsubsection{$Re=1$, $Wi=1$, $\varepsilon=0$, $\beta=1$, $Ma=0.05$, varying resolution $s$}
\begin{figure}
    \includegraphics[width=0.6\textwidth]{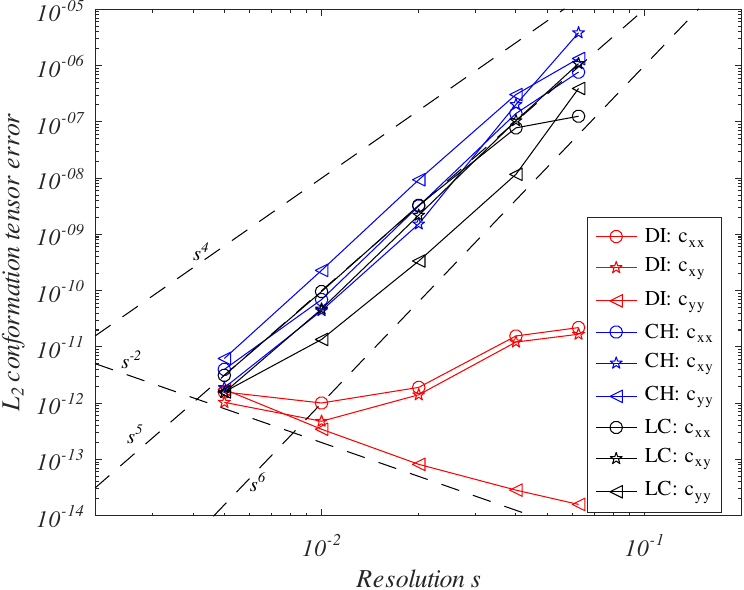}
    \caption{Variation of the Poiseuille flow steady state $L_{2}$ error norm of the conformation tensor components with resolution for parameters $Re=1$, $Wi=1$, $\varepsilon=0$, $\beta=1$ and $Ma=0.05$, for the three formulations: DI -red lines, CH - blue lines, LC - black lines. The annotations indicate the slopes of the dashed lines.\label{fig:p_conv_b1}}
\end{figure}

This selection of parameters decouples the momentum equation from the conformation tensor, and is a good validation of the accuracy of our discretisation. Figure~\ref{fig:p_conv_b1} shows the convergence with resolution of the $L_{2}$ norm of the conformation tensor components once a steady state has been reached. For direct integration of the conformation tensor, results are much more accurate. At these low $Re$, the time-step is limited by the viscous constraint, and so $\delta{t}\propto{s}^{2}$. This explains the increasing error from machine precision with resolution, clearly visible for $c_{yy}$ in particular - there is an accumulation of time-stepping errors of order $s^{-2}$. For LC and CH approaches, we see more typical convergence behaviour of order $5$ with resolution $s$. Indeed, the DI approach would also exhibit $5$th order convergence if the errors were not already so close to machine precision, and this is shown in subsequent results in the next section. 

The $5$th order convergence behaviour observed follows principally from the boundary conditions employed. Although the finite difference stencils used on the wall boundary are $4$th order, the dominant error in this Poiseuille flow arises mainly due to errors in the nonlinear advection terms, which are larger in the near-wall region, where the order of the LABFM discretisation is reduced from $m=8$ to $m=6$. The elements of $\bm{c}$ have quadratic and linear forms, and therefore for DI, their derivatives are reproduced exactly. Due to the Cholesky- and log- transformations, the elements of both $\bm{L}$ and $\bm{\Psi}$ have more complex forms, and first derivatives are accurate to order $m$, whilst second derivatives are accurate to order $m-1$. The order $m=6$ consistency near the walls, combined with the accumulation of time-stepping errors, results in the observed convergence rates of order $5$.

\subsubsection{$Re=1$, $Wi=1$, $\varepsilon=0$, $\beta=0.1$, varying $Ma$ and resolution $s$}

\begin{figure}
    \includegraphics[width=0.6\textwidth]{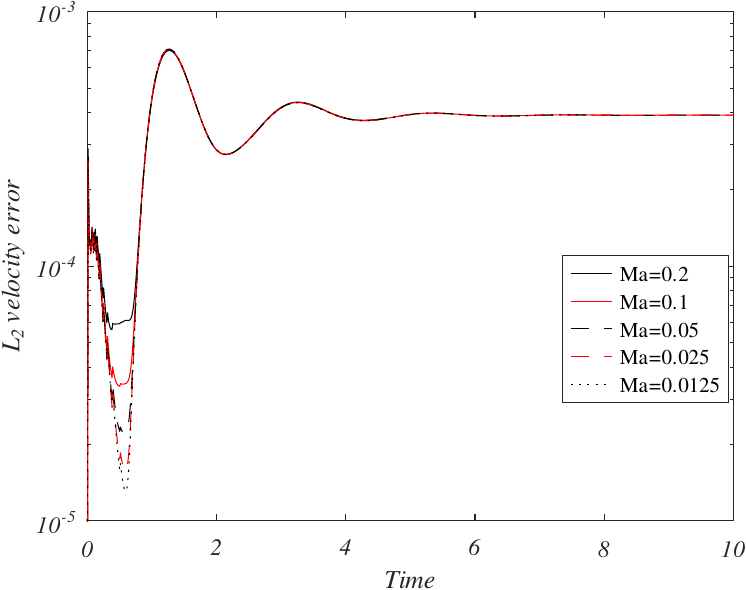}
    \caption{Poiseuille flow: Time evolution of the $L_{2}$ error in the velocity for several values of $Ma$, using log-conformation formulation (denoted LC), with a resolution of $s=1/50$.\label{fig:p_mach}}
\end{figure}

We now demonstrate the effect of varying Mach number on the solution. At steady state, the solution has uniform pressure and so offers a good test on the role of Mach number as compressibility should not affect the final solution. This is confirmed by Fig~\ref{fig:p_mach} showing the time evolution of the $L_{2}$ error in the velocity for several values of $Ma$ using LC and a resolution $s=1/50$. There are small error differences in the transient flow at early times, but the steady state is unaffected. Accordingly, for the remainder of this section, we use $Ma=0.05$. In~\cite{king_2022} we also showed the effects of changing $Ma$ but in the Newtonian setting by comparing to analytical solutions for Taylor Green vortices, and also found $Ma=0.05$ to be a suitable value. However, viscoelastic simulations are invariably more complex and some care does need to be taken at larger $Wi$, where the value of $Ma$ can affect stability of the simulation, particularly if $Ma$ is too small, and also at very low $Re$, where large viscous stresses can require smaller $Ma$ to ensure negligible compressibility. This behaviour is described further below.   

\begin{figure}
    \includegraphics[width=0.49\textwidth]{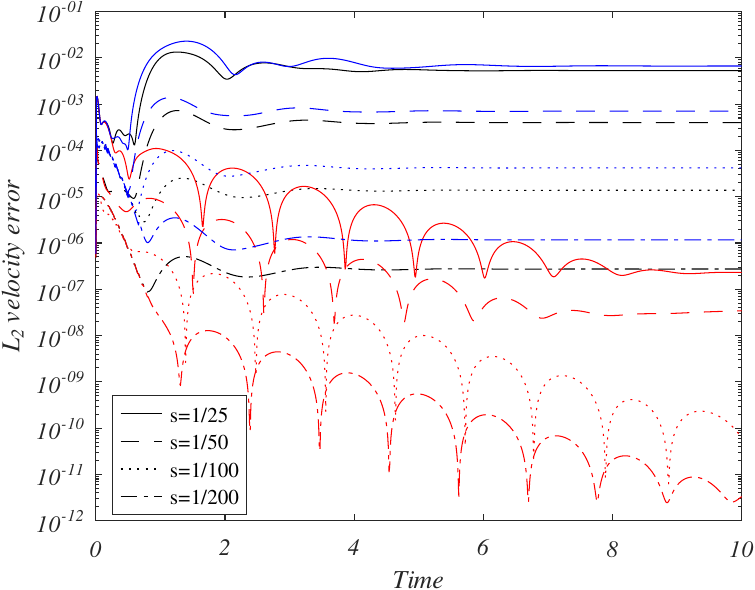}
    \includegraphics[width=0.49\textwidth]{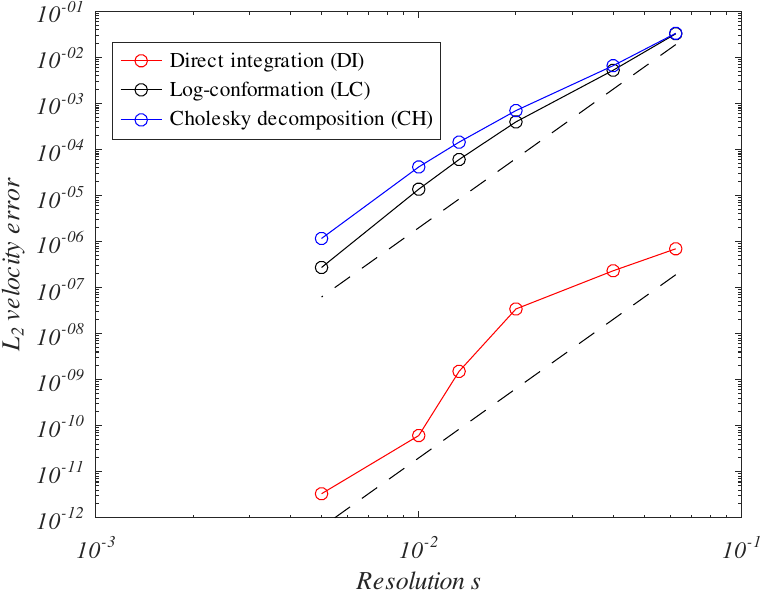}    
    \caption{Poiseuille flow. Left panel: Time evolution of the $L_{2}$ error in the velocity for all three formulations (DI - red lines, CH - blue lines, LC - black lines), for several resolutions $s$. Right panel: Convergence of the $L_{2}$ error in the velocity at time $t=10$ with resolution for all three formulations. The dashed lines show convergence rates of order $5$.\label{fig:p_conv}}    
\end{figure}

The left panel of Figure~\ref{fig:p_conv} shows variation of $L_{2}$ error in velocity with time for several resolutions, for direct integration of the conformation tensor (DI), Cholesky-decomposition (CH) and the log-conformation formulation (LC). The right panel of Figure~\ref{fig:p_conv} shows the variation of the $L_{2}$ error in velocity at $t=10$ with resolution $s$ for all three formulations.

LC, CH and DI all converge with approximately $5$th order. The DI approach is more accurate by several orders of magnitude. The initial oscillation in the error seen in the left panel of Figure~\ref{fig:p_conv} results from the interplay between the elastic stresses and small acoustic waves generated at start-up, with the size of these small amplitude oscillations decreasing with decreasing $Ma$. The larger error magnitude for DI here than in the previous case with $\beta=1$ arises because, although the steady solution is quadratic in $y$, the transient solution is not. As a result, the advective errors described in the previous subsection for LC and CH also occur for DI in the present case where $\beta\ne1$.

As indicated above, at lower $Ma$ the solution for all cases becomes less stable, with a gradual increase in error at late times, which appears to be very small (machine precision order) errors acting multiplicatively every time step. Indeed, for the resolution $s=1/200$, the time-step is $\delta{t}=7.5\times{10}^{-6}$ (dimensionless units), so a simulation up to $t=10$ requires a significant number of time-steps (more than $10^{6}$). Evidently, larger $Ma$ values introduce a degree of acoustic dissipation that limits such error growth and helps to stabilise the simulation. In this regard setting an upper limit of $Ma=0.05$ for simulations provides a good compromise between maintaining numerical stability and providing a good approximation to flow incompressibility. 

\subsubsection{$Re=1$, $\beta=0.1$, $\varepsilon=0$, $Ma=0.05$, $s=1/100$, varying $Wi$}

\begin{figure}
    \includegraphics[width=0.49\textwidth]{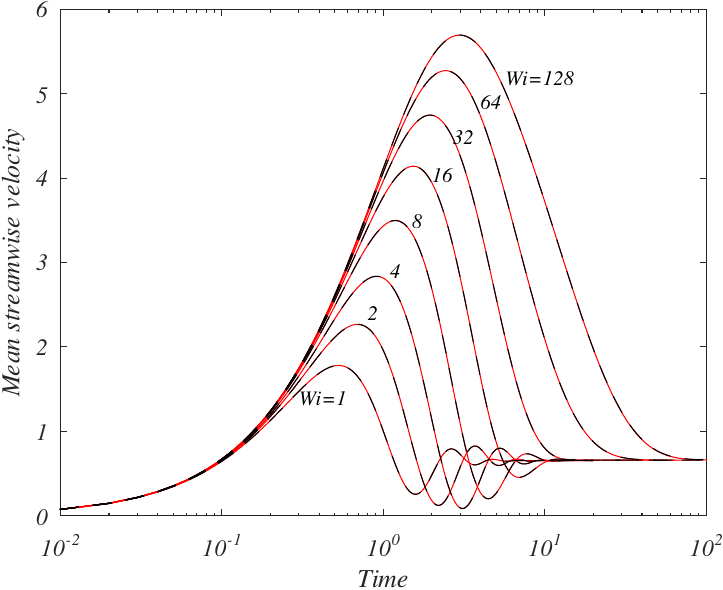}
    \includegraphics[width=0.49\textwidth]{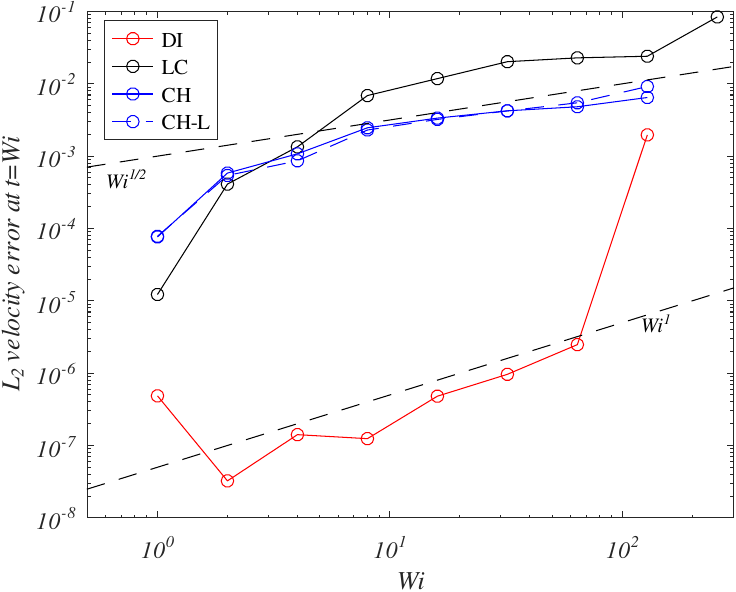}    
    \caption{Poiseuille flow. Left panel: Time evolution of the mean velocity for a range of $Wi$, obtained with the LC formulation. Red lines indicate the analytic solution, dashed black lines indicate the numerical results. Right panel: Variation with $Wi$ of the $L_{2}$ error in the velocity at dimensionless $t=Wi$, for all formulations. In all cases $s=1/100$. \label{fig:p_wi}}    
\end{figure}

In this subsection we vary the Weissenberg number, $Wi$, to determine the accuracy of the solution at different levels of elasticity and to explore the largest allowable $Wi$ for a stable simulation for this test case. The left panel of Figure~\ref{fig:p_wi} shows the time evolution of the mean velocity for a range of values of $Wi$ (dashed black lines) compared with the unsteady analytical solution (red lines). Results shown here were obtained with LC, but the results for DI and CH are indistinguishable on these axes. As can be seen, there is an excellent match for $t<100$ up to $Wi=128$. At this resolution, the CH and DI approaches break down above $Wi=128$, whilst LC is stable at $Wi=256$. Beyond these $Wi$ values, at this resolution, the three schemes fail. All three formulations are capable of reaching higher $Wi$ if the resolution is increased ($s$ reduced) further as key terms in the governing equations are better resolved.

In particular, the errors in evaluating advection terms are larger with increasing $Wi$, resulting in deviation of $c_{yy}$ from unity for LC and CH \hlc{(and CH-L)}, with this deviation increasing with $Wi$ and decreasing with larger $s$. Typically errors of order $10^{-2}$ when $Wi=16$ and $s=1/100$ are seen in velocity and stress profiles for LC and CH\hlc{/CH-L} approaches, with DI errors smaller by orders of magnitude (but still growing with increasing $Wi$). This behaviour can be seen in the right panel of Figure~\ref{fig:p_wi} which shows the $L_{2}$ error in the velocity at (dimensionless) $t=Wi$ for a range of $Wi$, for all three formulations. The orders of error growth with $Wi$ are indicated by the dashed lines, and whilst DI has lower overall error, the rate of error growth with increasing $Wi$ is larger than the LC and CH formulations. 

The difference in growth of errors with $Wi$ between DI and LC and CH can be attributed to differences in the solution profiles across the channel. As discussed earlier, whilst in the steady state the cross-channel profiles of $c_{xx}$ and $c_{xy}$ are quadratic and linear (respectively) in $y$, the log- and Cholesky- transformations of $\bm{c}$ result in the components of $\bm{\Psi}$ and $\bm{L}$ having profiles with more complex structure. In particular, whilst the profiles of the components of $\bm{\Psi}$ and $\bm{L}$ are nearly linear and quadratic near the channel walls, they have greater curvature in the channel centre (where the stress is zero), and this curvature increases for increasing $Wi$. As such, at larger $Wi$, the dominant errors for LC and CH occur near the channel centre, whilst for DI they are more uniform across the domain (with the only variation being near the walls, where $m$ is reduced towards $m=4$). \hlc{Finally, we note that the growth of errors with $Wi$ is very similar for both CH and CH-L formulations, as it is the non-exact (but still $\mathcal{O}\left(s^{m}\right)$) advection of the non-linear profiles near zero-stress points which dominates in this case.}

\subsection{Periodic array of cylinders}

This flow case provides a test of the method in a non-trivial geometry and allows us to assess the performance of the different formulations (LC, CH and DI) for non-parallel flows. The periodic cylinders case simulated follows that of~\cite{king_2021} and is based on~\cite{vq_ellero_2012}. The domain is rectangular with dimension $6R\times{4}R$, with a cylinder of radius $R$ located in the centre. At the upper and lower boundaries, and the cylinder surface, no-slip wall boundary conditions are imposed, whilst the domain is periodic in the streamwise direction. The flow is driven by a body force, the magnitude of which is set by a PID controller such that the mean velocity magnitude (averaged over the domain) is unity. We take the cylinder radius $R$ as the characteristic length-scale for non-dimensionalisation.

In all cases we set $Re=2.4\times{10}^{-2}$, $\beta=0.59$ and $\varepsilon=0$, to match those parameters used in~\cite{vq_ellero_2012,king_2021}. For this non-parallel flow case at low $Re$, the magnitude of the body force required to drive the flow is larger than the previous cases, and a smaller value of $Ma$ is required to ensure density variations remain small. We set $Ma=10^{-3}$, which results in density variations of less than $0.5\%$. For this test we discretise the domain with a uniform resolution $s_{a}=s$ for all nodes $a$, allowing comparison with the aforementioned previous works. 

\begin{figure}
    \includegraphics[width=0.49\textwidth]{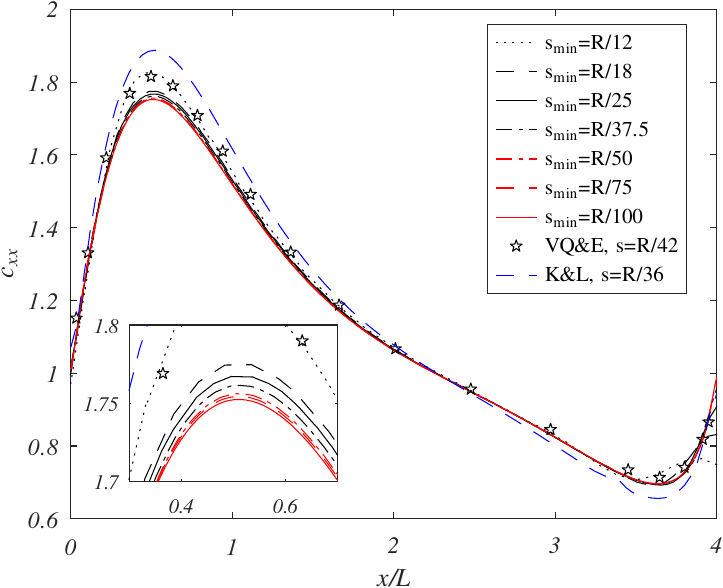}
    \includegraphics[width=0.49\textwidth]{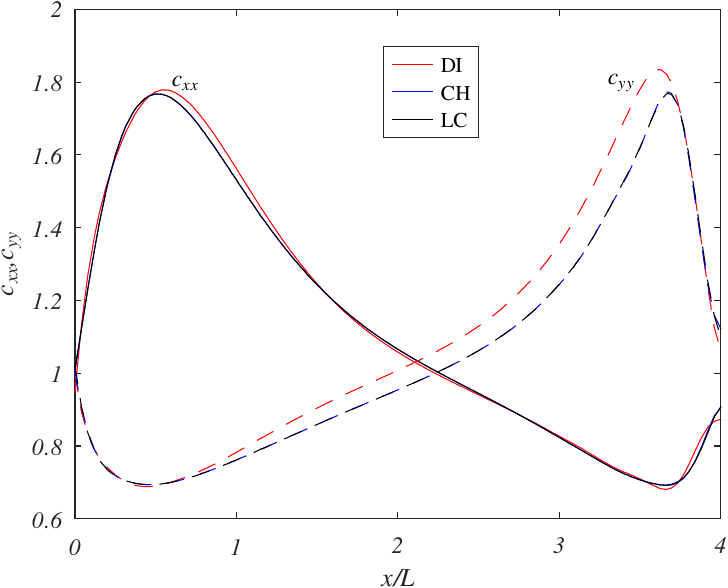}    
    \caption{Profiles of the conformation tensor components along the channel centreline for $Wi=0.2$ at steady state (20 dimensionless time units). Left panel: $c_{xx}$ for a range of resolutions, using the log-conformation formulation. Results taken from SPH simulations are shown with black stars (~\cite{vq_ellero_2012}) and a dashed blue line (~\cite{king_2021}). Right panel: $c_{xx}$ (solid lines) and $c_{yy}$ (dashed lines) for a resolution of $s=R/25$ for the three different formulations.\label{fig:cylwi0p2}}
\end{figure}

We first set $Wi=0.2$ and assess the accuracy of the method using the LC formulation. The left panel of Figure~\ref{fig:cylwi0p2} shows the profile $c_{xx}$ along the channel centreline for a range of resolutions. We see clearly see convergence in the LABFM solution (inset). The results are compared with SPH data from~\cite{vq_ellero_2012,king_2021}, with good agreement shown despite SPH being formally low order. Indeed, both schemes in~\cite{vq_ellero_2012,king_2021} benefit in not having to compute the advection term by their nature of being Lagrangian methods, removing a key term for error growth in the LABFM method. Furthermore, the formulation of~\cite{vq_ellero_2012} is constructed in the GENERIC framework: the symmetries of the conservation laws are matched by the discretised formulation in a thermodynamically consistent way, providing benefits for longer term dynamics and global conservation. 

The right panel of Figure~\ref{fig:cylwi0p2} shows the profiles along the channel centreline of $c_{xx}$ and $c_{yy}$ for a fixed resolution of $s=R/25$, for the three formulations DI (red), CH (blue) and LC (black). The LC and CH formulations are almost indistinguishable, but the values from the DI approach (and $c_{yy}$ in particular) deviate slightly. 

\begin{figure}
    \includegraphics[width=0.99\textwidth]{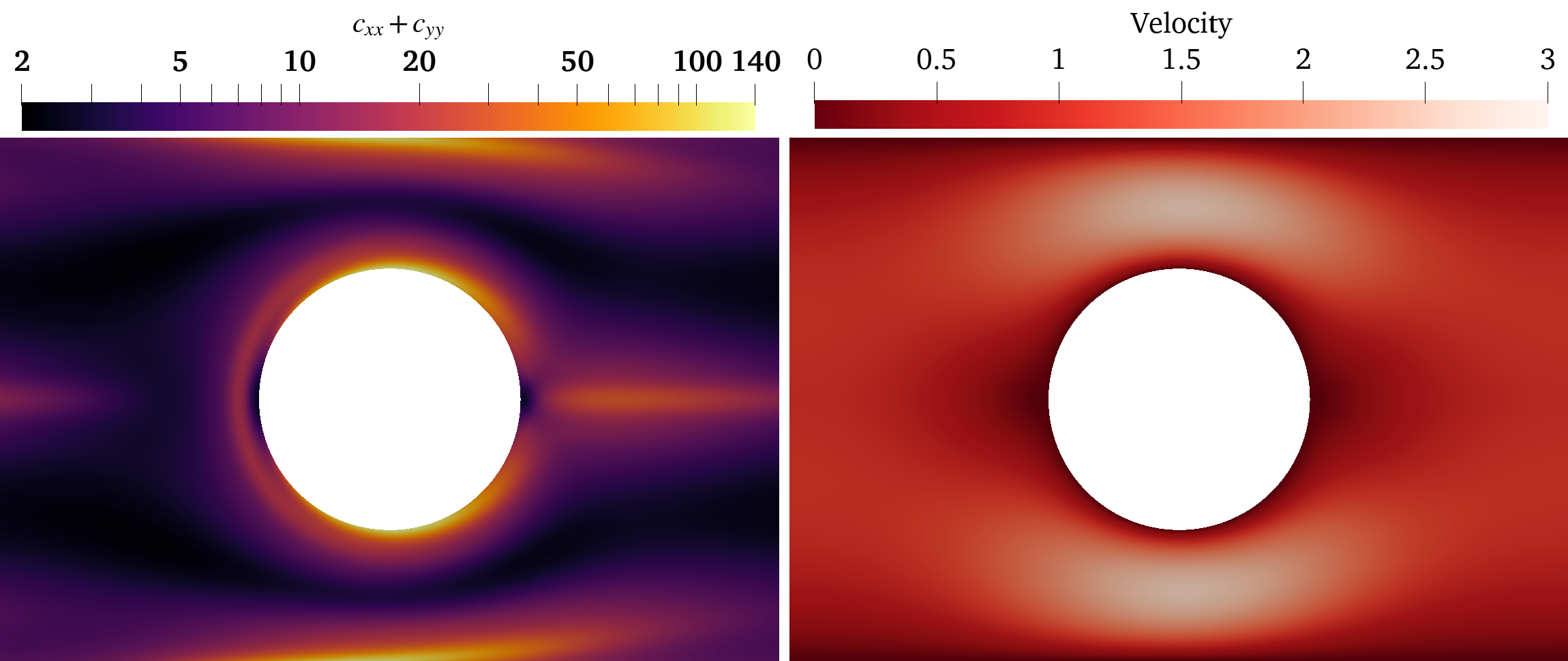}
    \caption{Isocontours of conformation tensor trace (left) and velocity magnitude (right) for the periodic array of cylinders with $Wi=0.8$ and $s_{min}=R/50$, obtained using the log-conformation formulation.\label{fig:cyl}}
\end{figure}

\begin{figure}
    \includegraphics[width=0.49\textwidth]{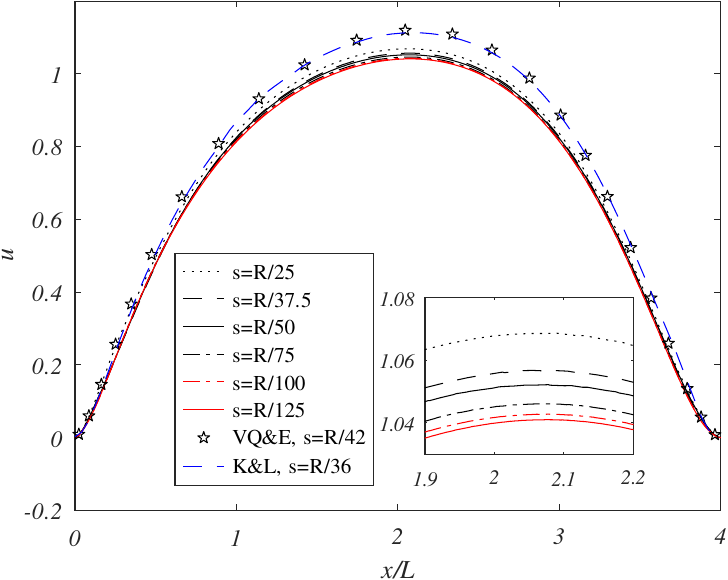}
    \includegraphics[width=0.49\textwidth]{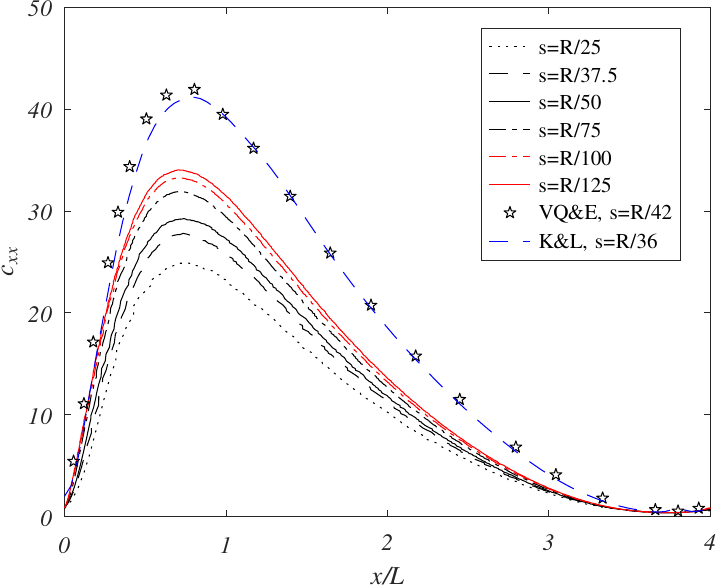}    
        \caption{Profiles along the channel centre-line of the velocity (left panel) and conformation tensor component $c_{xx}$ (right panel) \hlb{at $Wi=0.8$}, for a range of resolutions (red and black lines), compared with the SPH results of~\cite{vq_ellero_2012} (black stars) and~\cite{king_2021} (dashed blue lines).\label{fig:cylwi0p8}}    
\end{figure}

We next increase the degree of elasticity, setting $Wi=0.8$. \hla{Figure~\mbox{\ref{fig:cyl}} shows the conformation tensor trace and velocity magnitude fields in this case, with $s_{min}=R/50$, obtained with LC.} The problem becomes more numerically challenging now, as the infinite polymer extensibility of the Oldroyd B model results in a singularity in the stress field in the cylinder wake. It was found by~\cite{alves_2001} through numerical experiments that for a cylinder in a channel (no periodicity assumed), the solution was divergent for $Wi\ge0.7$, whilst a similar result was obtained by~\cite{bajaj_2008}, who obtained exact solutions for the wake centreline stress in the ultra-dilute case. Non-convergence with resolution for $Wi=0.8$ was also observed in~\cite{vq_ellero_2012}. 
For the direct integration formulation, a catastrophic instability occurs early in the simulation at all resolutions tested. This is due to errors in the advection of the elements of $\bm{c}$, which result in a loss of positive definiteness, leading to non-physical results. At all resolutions studied, CH and LC are in close agreement. In the remainder of this section, the results presented are obtained using the LC formulation. Figure~\ref{fig:cylwi0p8} shows the profiles of velocity (left) and conformation tensor component $c_{xx}$ (right) along the channel centre line for a range of resolutions, alongside results from SPH simulations of~\cite{vq_ellero_2012} (black stars) and~\cite{king_2021} (dashed blue lines). 

Firstly, it is clear that the profile of the stress in the cylinder wake is diverging with resolution refinement as expected, and we note that the maximum value of $c_{xx}$ in the cylinder wake scales linearly with $R/s$. Note, that although the stress field is divergent, the velocity field is converging with increasing resolution (inset of the left panel of Figure~\ref{fig:cylwi0p8}). Secondly, there are clear discrepancies between the present results and the results of~\cite{king_2021,vq_ellero_2012} using SPH. As described above, there are several contributing factors behind differences observed in Figure~\ref{fig:cylwi0p8} between the SPH simulation results of~\cite{king_2021,vq_ellero_2012} and the present work - not least of which being that SPH is formally low order, with such differences in method accuracy exacerbated in a parameter regime with a divergent stress field incorporating steep gradients. \hlb{However, the exact cause of the discrepancy is not clear, and we also note that in the cited SPH results, the method is Lagrangian, and the non-linear advection terms are implicitly included in the temporal evolution of the particle positions.}

\subsection{Representative porous geometry}

\begin{figure}
    \includegraphics[width=0.99\textwidth]{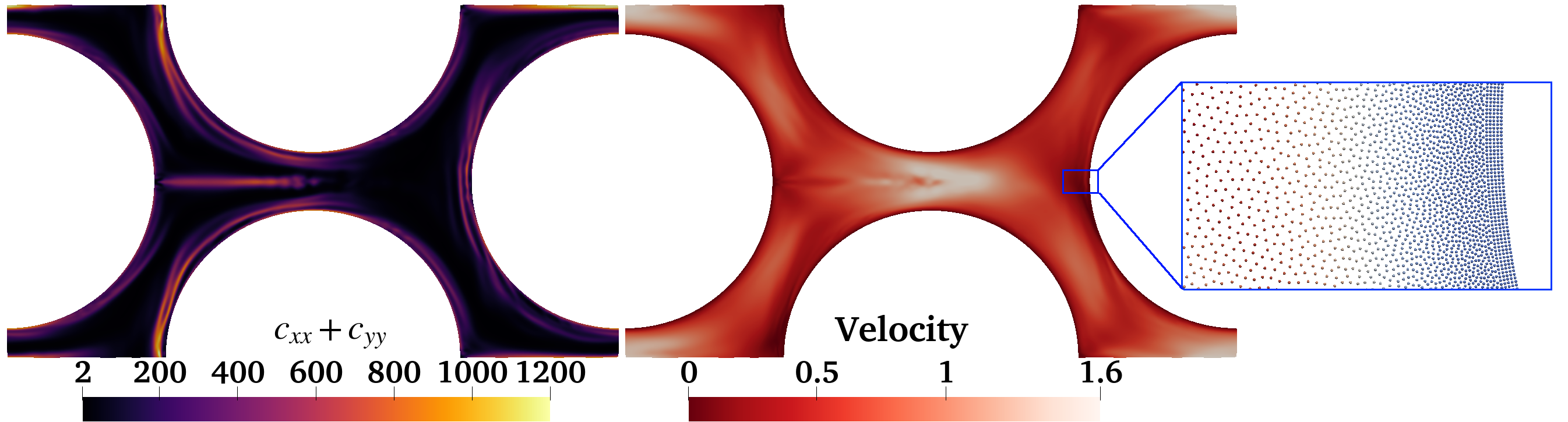}
        \caption{Isocontours of conformation tensor trace (left) and velocity magnitude (right) for the porous geometry with $Wi=1$ and $s_{min}=D/600$, simulated using LC. \hlb{The inset on the right shows the node distribution near one of the cylinders.}\label{fig:porous1}}    
\end{figure}

We next consider a repeating unit of a representative porous geometry, consisting of cylinders with diameter $D$ and spacing $S$. \hlc{Several authors have studied similar configurations, with Lattice-Boltzmann methods~\mbox{\cite{dzanic_2023,gillissen_2013}} and finite volume methods~\mbox{\cite{de_2017,kumar_2021}}. All these works study the flow at negligible $Re$, whilst we use $Re=1$. Whilst there are differences in the exact geometries between these works, they all contain the same essential features, allowing for qualitative comparisons to be made.} The computational domain has size $\sqrt{3}S\times{S}$, and is periodic in both directions. A cylinder of diameter $D=S/1.2$ is centred on the midpoint of each boundary. The domain therefore represents a minimal repeating unit of a hexagonal lattice of cylinders. The geometry can be seen in Figure~\ref{fig:porous1}. The system is non-dimensionalised by the cylinder diameter $D$ and the mean velocity magnitude $U$. The flow is driven by a body force in the $x$ direction, which is set by a PID controller to track $U=1$. The domain is discretised with a non-uniform resolution $s_{min}$ at the cylinder surfaces, and $s$ increases smoothly away from the cylinders to $s_{max}=3s_{min}$ at distances greater than $25s_{max}$ from the cylinders. \hlb{The node distribution near the cylinder is shown in the inset on the right of Figure~\mbox{\ref{fig:porous1}}.} With the finest flow structures located near the cylinder walls, the accuracy of the simulations is largely controlled by $s_{min}$, which we use to characterise the resolution of each simulation. 

In all cases, we set $Re=1$, $\beta=0.5$, $\varepsilon=10^{-3}$, $Ma=10^{-2}$. We vary $Wi$ and the resolution. The inclusion of non-zero $\varepsilon$ (thus representing a PTT fluid rather than an Oldroyd B fluid) avoids the singularity present in the previous test case.
We note that the values of $Wi$ studied are small, and based on $U$ and $D$. An effective Weissenberg number $Wi_{eff}$ for the flow within the pore space may be a more pertinent measure, and could be defined based on the pore size $S-D=D/5$, giving $Wi_{eff}=5Wi$. 

\subsubsection{Effect of formulation and resolution at fixed $Wi=1$}

In the first instance we run the simulation for all three formulations. For DI, the simulation quickly becomes unstable, as the thin regions where $c_{yy}$ are large just upstream of each cylinder (see left panel of Figure~\ref{fig:porous1}) cannot be accurately advected. Local oscillations occur, resulting in negative values of $c_{yy}$, and loss of positive definiteness of the conformation tensor. For CH, the simulation exhibits numerical artefacts at lower resolutions than LC. It appears that at higher resolutions, CH is capable of handling this problem, but LC can achieve accurate solutions at lower resolution, and hence lower cost. 

\begin{figure}
    \includegraphics[width=0.49\textwidth]{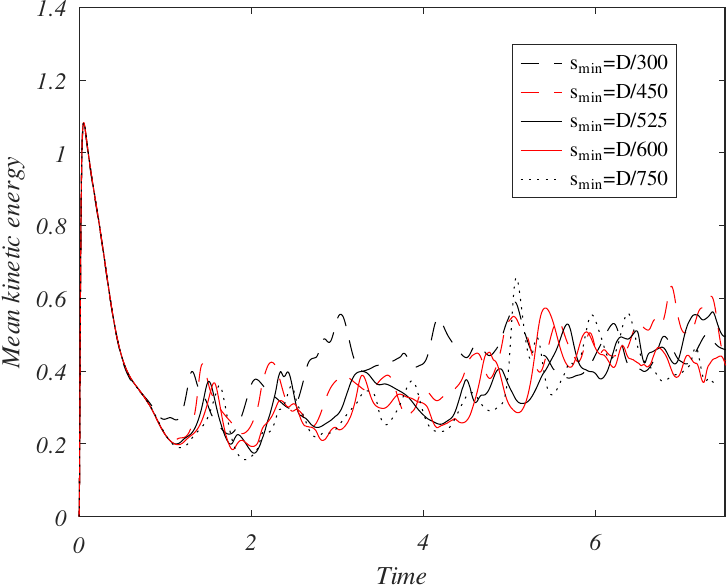}
    \includegraphics[width=0.49\textwidth]{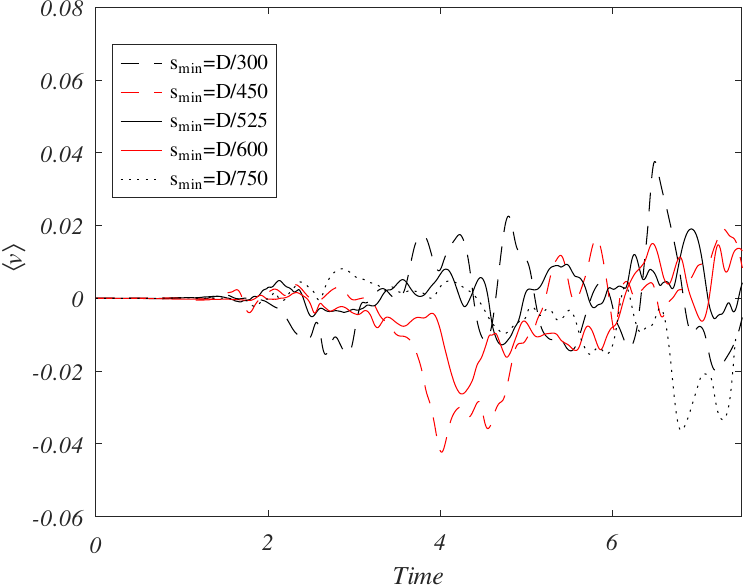}    
        \caption{The time evolution of the volume averaged value kinetic energy (left panel) and transverse velocity $\left\langle{v}\right\rangle$ (right panel), for a range of resolutions.\label{fig:porous_res}}    
\end{figure}

With LC identified as the best formulation for this problem, we focus in more detail on the effect of resolution. In all cases hereafter, we use the LC formulation. Figure~\ref{fig:porous_res} shows the time variation of volume averaged kinetic energy and transverse velocity for resolutions $s_{min}/D\in\left[1/300,1/450,1/525,1/600,1/750\right]$, when using the log-conformation formulation. For resolutions finer than $s_{min}=D/525$ the kinetic energy is approximately converged, and the global drag in the system (as measured by the body force required to drive the flow) is converged to within $2.4\%$. The entire system has a  chaotic/sensitive dependence on initial conditions, and hence we don't see convergence in the exact trajectory of these global statistics. This behaviour is especially obvious in the right panel of Figure~\ref{fig:porous_res} showing the symmetry breaking given the considerable variation in the volume averaged transverse velocity, $\left\langle{v}\right\rangle$.

A particular computational challenge is that for low $Re$, we require very small time-steps due to the viscous time-step constraint, with the finest resolution $s_{min}=D/750$ requiring $>10^{7}$ time steps to simulate $8$ dimensionless time units. Conversely, at higher $Re$ we need exceptionally fine resolution to stably resolve the steep stress gradients and transients leading to the onset of elastic instability. Indeed, whilst high-order discretisations are invaluable for this problem, there is significant benefit to be had from variable and potentially adaptive resolution (in addition to high-order interpolants) for simulations of elastic instabilities. A fully implicit method utilising the present high-order interpolants and discretisation scheme would permit larger time-steps, and enable these simulations at reduced costs. Such an approach is an avenue we are interested in pursuing for future work.

\subsubsection{Symmetry breaking with increasing $Wi$}
As a precursor to the complete study and direct numerical simulation of elastic instability in this complex geometry, we consider in more detail the case of symmetry breaking in the flow with increasing $Wi$ up to $Wi=1$. \hlc{Note that whilst we show and quantify symmetry breaking, this is a preliminary study and our main focus is on the numerical method.} All results in this section are obtained using the LC approach with $s_{min}=D/600$. The Weissenberg numbers conisdered are $Wi\in\left[0,0.1,0.25,0.3,0.35,0.4,0.5,0.75,1\right]$. Beyond these values (approximately $Wi=1.5$) we expect transition to three-dimensional flow, as reported in ~\cite{grilli_2013} (for example), and hence extension to 3D simulations remains an area for future work.

\begin{figure}
    \includegraphics[width=0.49\textwidth]{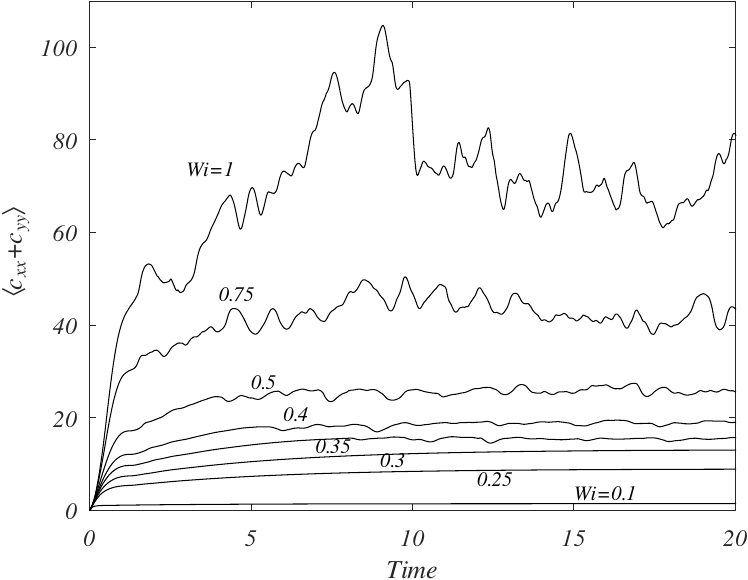}
    \includegraphics[width=0.49\textwidth]{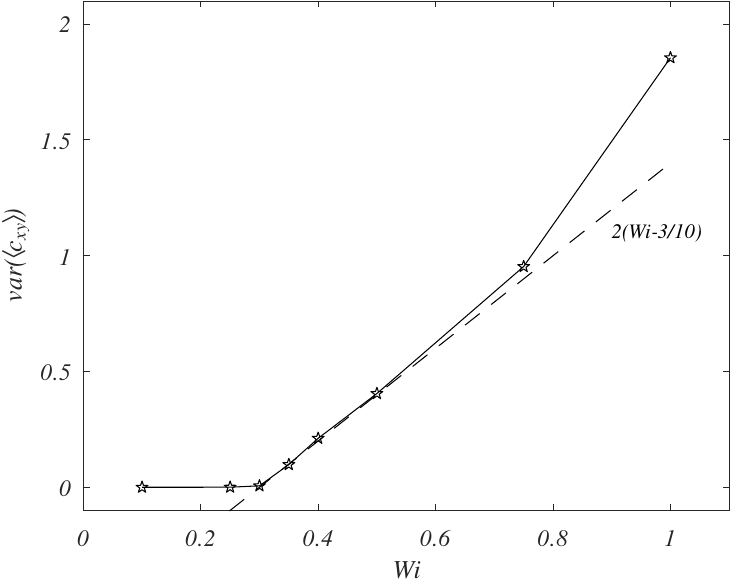}    
        \caption{Left panel: The time evolution of the volume averaged value of the conformation tensor trace $c_{xx}+c_{yy}$. Right panel: The variation with $Wi$ of the variance of the volume averaged value of $c_{xy}$. The dashed line illustrates a linear dependence on $Wi$.\label{fig:symmbreaking1}}    
\end{figure}

We define the instantaneous volume averaged conformation tensor elements as \begin{equation}\left\langle{c}_{ij}\left(t\right)\right\rangle=\displaystyle\int_{V}c_{ij}\left(t\right)dV,\end{equation}
which corresponds to the volume averaged trace if $j=i$, and the volume average of $c_{xy}$ if $j\ne{i}$. The left panel of Figure~\ref{fig:symmbreaking1} shows the time evolution of the volume averaged conformation tensor trace. As expected, fluctuations of increasing magnitude are seen with increasing $Wi$, but with values of the volume averaged conformation tensor trace levelling out (on average) with time, indicative of a statistically steady state. 

We evaluate the variance of $\left\langle{c}_{xy}\right\rangle$ once the statistically steady state has been reached, over the interval $t\in\left[10,20\right]$. This is a proxy for the measure of asymmetry in the polymeric deformation field. The right panel of Figure~\ref{fig:symmbreaking1} shows the dependence of $var\left\langle{c}_{xy}\right\rangle$ on $Wi$.  We see at small $Wi$ where the flow is steady, the variation is negligible. At $Wi\ge0.3$ the flow is unsteady and symmetry is broken, with the extent of the symmetry breaking having a linear dependence on $Wi$ (dashed line) with slope $2$. Note that for $Wi>0.75$ this relation ceases, likely as the flow enters a different, more elastic, regime. 

\begin{figure}
    \includegraphics[width=0.99\textwidth]{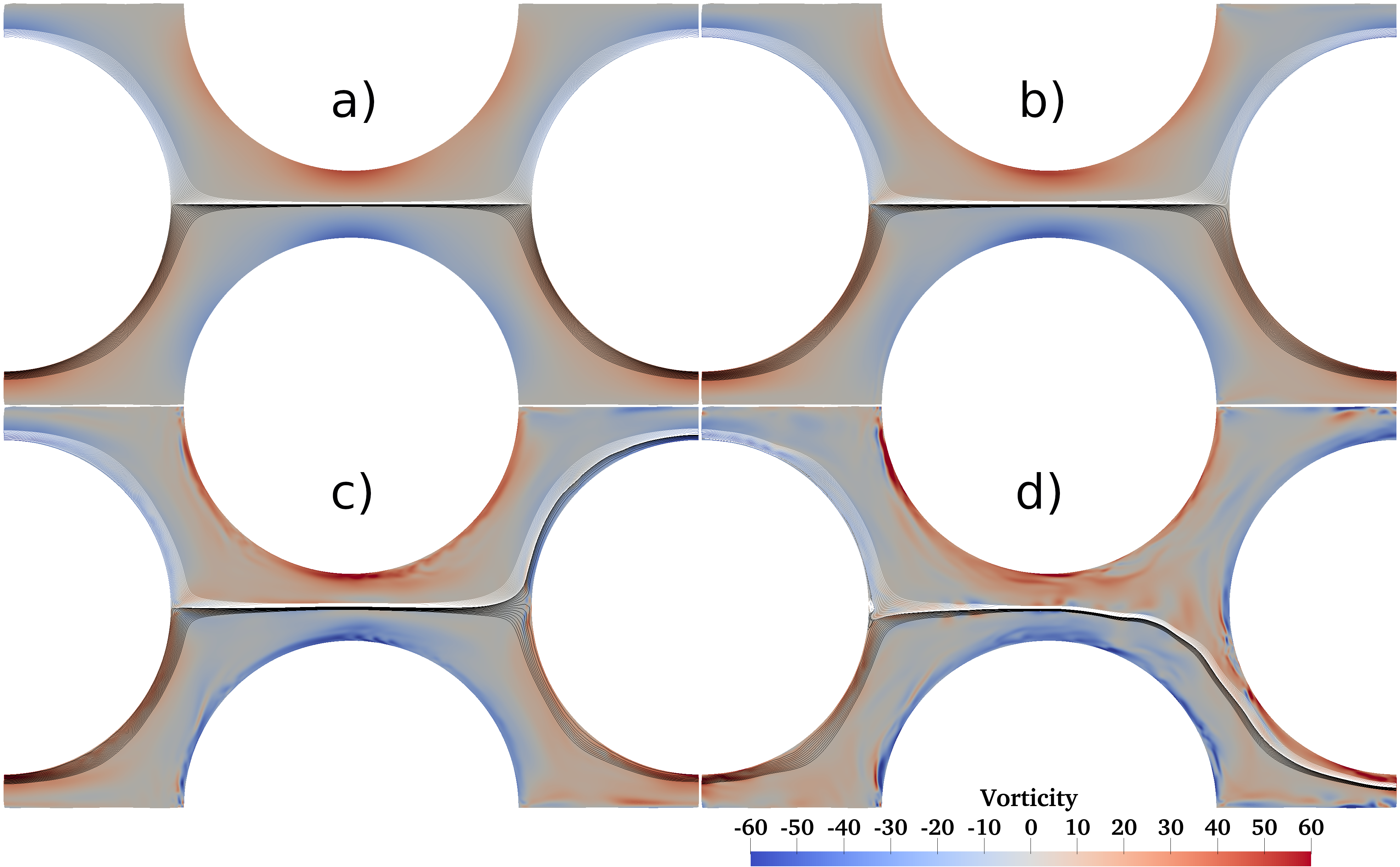}    
        \caption{Isocontours of vorticity (red-blue), with streamlines superimposed, showing the symmetry breaking with increasing $Wi$. Streamlines originating in the upper half of the domain are coloured white, and those originating in the lower half are coloured black. Panels: a) $Wi=0.0$, b) $Wi=0.25$, c) $Wi=0.5$, and d) $Wi=1$.\label{fig:symmbreaking2}}    
\end{figure}

\begin{figure}
    \includegraphics[width=0.99\textwidth]{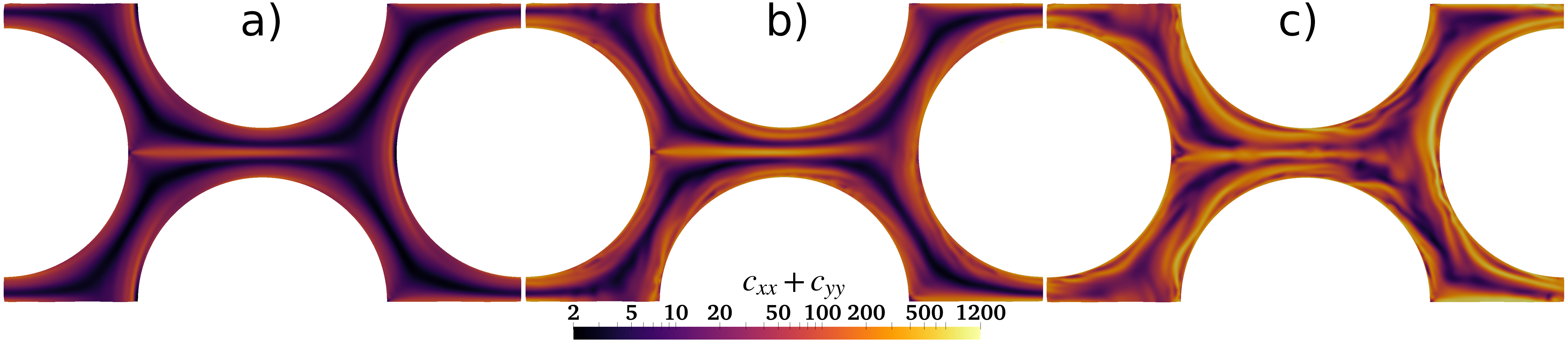}   
        \caption{Isocontours of conformation tensor trace for a) $Wi=0.25$, b) $Wi=0.5$, and c) $Wi=1$,\label{fig:symmbreaking3}}    
\end{figure}

Figure~\ref{fig:symmbreaking2} shows isocontours of the vorticity field (red-blue) with streamlines showing flow crossing from the upper to lower halves of the domain at increasing $Wi$. Note that between $Wi=0$ and $Wi=0.25$ (panels a) and b) respectively) the vorticity field develops a streamwise asymmetry\hlc{ as expected, and observed in a similar configuration by~\mbox{\cite{de_2017}}}. By $Wi=0.5$ the instability has developed and the symmetry is broken in the transverse direction, as shown by streamlines crossing the domain centreline. Similarly, Figure~\ref{fig:symmbreaking3} shows isocontours of the conformation tensor trace $c_{xx}+c_{yy}$ for a) $Wi=0.25$, b) $Wi=0.5$, and c) $Wi=1$. As above, by $Wi=0.25$ the field has developed a streamwise asymmetry, which clearly breaks in the transverse direction by $Wi=0.5$. Beyond this ($Wi=1$), symmetry breaking in the flow is clear with unsteady elastic flow structures larger in magnitude - a precursor stage before the flow evolves fully 3D flow structures. \hlc{The unsteady thin near-wall structures are qualitatively similar to those found in\mbox{\cite{dzanic_2023}}, who studied a similar geometry but in the creeping flow regime.} 

\section{Conclusions}\label{sec:conc}

In this work a new high-order meshless method for the solution of viscoelastic flow in two-dimensional, non-trivial geometries has been presented. Three different approaches to treating the viscoelastic stresses are considered for assessment in this new high-order meshless framework - direct integration, Cholesky decomposition, and the log-conformation formulation. Direct integration provides notably more accurate solutions for parallel flows but the log-conformation approach provides enhanced stability across all test cases considered. Highly accurate results can be obtained with convergence up to $9^{th}$ order, depending on the test case. \hla{For parallel flows, the attainable Weissenberg numbers are large, }up to $Wi=128$. \hla{For non-trivial geometries, the attainable Weissenberg numbers are more moderate, up to $\mathcal{O}\left(1\right)$, but we find that the limiting factor is the requirement to resolve the increasingly fine flow features present with increasing $Wi$, suggesting tha our method can handle higher $Wi$ given sufficient resolutions.} The meshless nature of the method enables non-trivial geometries to be discretised straightforwardly, with variable resolution easily included. Accordingly, an initial study of a symmetry breaking elastic instability at moderate $Wi$ is considered in a non-trivial representative porous media geometry. The results are promising and demonstrate the potential of this method for the high-fidelity study of fully 3-D elastic instabilities in realistic, industrially relevant geometries in the longer term. \hlb{The explicit nature of the present formulation renders simulations in the limit of vanishing $Re$ impractical, and thus the method is well suited to inertial flows. An implicit formulation would allow us to simulate flows with negligible inertia, and explore purely elastic instabilities in complex geometries.} These are the main goals of our future work, with any 3-D method also requiring adaptivity of resolution, both in polynomial reconstruction and spatial resolution, to enable the capture of thin, unsteady elastic flow structures in a computationally efficient manner.

\begin{acknowledgments}
JK is funded by the Royal Society via a University Research Fellowship (URF\textbackslash R1\textbackslash 221290). We would like to acknowledge the assistance given by Research IT and the use of the Computational Shared Facility at the University of Manchester. 
\end{acknowledgments}

\bibliography{jrckbib}

\end{document}